\makeatletter
\def\setcaptype#1{\def\@captype{#1}}
\makeatother
\documentstyle[11pt]{article}

\title{A model for the $Q^2$ dependence of polarized structure
functions.}
\author{B.~Ehrnsperger and A.~Sch\"afer
\\
Institut f\"ur Theoretische Physik, J.~W.~Goethe
Universit\"at Frankfurt,
\\
Postfach~11~19~32, D-60054~Frankfurt am Main, Germany
\\
UFTP 370/1994, hep-ph/9505306, to appear in PRD
}
\begin{document}
\maketitle
\begin{abstract}
We present an update of a phenomenological model for the spin dependent
structure functions $g_1(x,Q^2)$ of the proton and neutron.
This model is based on a broken SU(6) wavefunction
parametrized by the unpolarized structure functions.
The two free parameters of the model are choosen to fulfill
the Bjorken and Ellis--Jaffe sum rules.
The model respects isospin symmetry and has zero strange sea
polarization.
Using new values for $F/D$ from hyperon beta decay the resulting $Q^2$
dependent asymmetries $A_1$ are in perfect agreement with the existing data.
Therefore we do not see any evidence for a ``spin crisis''.
With two choices for $g_2$ the $Q^2$ dependence of $A_1(x,Q^2)$ and
$A_2(x,Q^2)\sqrt{Q^2}/M$ is predicted and shown to be small for both cases.
\end{abstract}
\\ \\
\section{Introduction \label{sec1}}
The first measurements of $g_1^p$ \cite{E130} suggested very
surprisingly that the quarks
carry much less of the protons spin than what was called the
``naive expectation''.
Subsequent experiments \cite{JAs88,PAn93,BAd93,DAd94,KAb95}
and theoretical papers stimulated
a lot of discussion concerning this so called ``spin crisis'',
much of which centered around the correctness of the
Ellis--Jaffe \cite{JEl74}
and the Bjorken \cite{JBj66} sum rule.
It became clear over the years that knowledge of the precise
$Q^2$--dependence is crucial for attaining unambiguous
results.
Luckily the finite $Q^2$ corrections to the theoretical predictions for the
sum rules get slowly under control \cite{SLa91,ESt94}.

To derive moments of the structure functions from
experiment it is
necessary to find the right extrapolation of the structure functions to small
$x$ for fixed $Q^2$.
As low $x$ experimentally corresponds to low $Q^2$
it is very difficult to treat this systematic effect correct
other than by much improved measurements over a much wider kinematic
range.
At present in all experiments the asymmerty $A_1$ is taken to be
$Q^2$ independent.
We only know of one previous data analysis \cite{GAl93,RBa95} that takes
the $Q^2$ dependence into account, however the authors of this
analysis use an approximation for the numerical evolution of the $Q^2$
dependence of $g_1$ which differs for low $Q^2$
from our results.

In studing the $Q^2$ dependence of the measured asymmetry $A_1$
(c.f. eq. (\ref{eq39})) we proceed in two steps:
In section \ref{sec2} we present a modification of the Carlitz Kaur model
\cite{RCa77,ASc88}
to get $g_1^p(x,Q_0^2)$, $g_1^n(x,Q_0^2)$ and a quantity
$\Delta \tilde{G}(x,Q_0^2)$, which is a lower bound for the real
$\Delta G(x,Q_0^2)$
for a fixed virtuality, e.~g.~ $Q_0^2 = 10$ GeV$^2$.
All parametres of the model
are fixed by the unpolarized structure functions, the Bjorken
\cite{JBj66} and the Ellis--Jaffe \cite{JEl74} sum rule.

In section \ref{sec2a} we present a modification of our model
that does not show the leading particle picture.
We show that it is able to fit the data as well.

In section \ref{sec3} the structure functions are evolved
in $Q^2$ using first order GLAP equations \cite{GAl77}.
Resulting Asymmetries $A_1^{p/n}$
are compared with data and predictions are made for $A_2^{p/n}(x,Q^2)$.

Section \ref{sec4} gives a summary and conclusions.
\section{A model for $g_1^p(x,Q_0^2)$, $g_1^n(x,Q_0^2)$ and
$\Delta \tilde{G}(x,Q_0^2)$ at $Q^2_0$ \label{sec2}}
The model which we present tries to preserve $SU(6)$
symmetry as far as possible.
However it is  known that SU(6) symmetry
is broken in (at least) two aspects.
First, the $u(x)$ and $d(x)$ distribution functions for the
nucleon are different,
and second, there is a  mass splitting between the nucleon ground state
and the delta resonances.

An $x$ dependent SU(6) symmetric wavefunction for the proton \cite{RCa77},
which describes the structure of the valence quarks is:
\begin{eqnarray} \label{eq1}
\big| p(x)\uparrow \big> &=& \hat{u}^\dagger (x) \uparrow
\big|  I=0,I_3=0,S=0,S_3=0  \big>  \sqrt{{\cal A}_0(x)/x}
\nonumber \\
         && + \left( \sqrt{2/3} \; \hat{d}^\dagger (x) \big| I=1,I_3=1
\big>  -
            \sqrt{1/3} \; \hat{u}^\dagger (x) \big| I=1,I_3=0  \big>  \right)
\\
        &&  \times \left(\sqrt{2/3} \; \downarrow \big| S=1,S_3=1 \big>
          - \sqrt{1/3} \; \uparrow \big| S=1,S_3=0 \big>  \right)
           \sqrt{{\cal A}_1(x)/x} \; .
\nonumber
\end{eqnarray}
In this picture the proton state $\big| p(x)\uparrow \big> $ consists of one
active quark (parton) carrying the momentum fraction $x$ and two spectator
quarks coupled to the indicated quantum numbers.
As usual $\hat{q}^\dagger (x)$ denotes a creation operator of quark type q
with momentum fraction $x$,
$q_v(x)$ denotes a valence quark distribution function of the proton:
\begin{equation}
q_v(x) := q_v^p(x) = \big< p(x)\uparrow \big| \hat{q}^\dagger (x)\uparrow
                 \hat{q}(x)\uparrow
             \;+\; \hat{q}^\dagger (x)\downarrow \hat{q}(x)\downarrow
        \big| p(x)\uparrow \big>
\end{equation}
an similar for $\Delta q_v(x)$:
\begin{equation}
\Delta q(x) = \Delta q_v(x) := \Delta q_v^p(x) = \big< p(x)\uparrow \big|
\hat{q}^\dagger (x)\uparrow
\hat{q}(x)\uparrow
             \; - \; \hat{q}^\dagger (x)\downarrow \hat{q}(x)\downarrow
        \big| p(x)\uparrow  \big>
\end{equation}
In the following we assume the validity of isospin symmetry
i.e.:
\begin{equation}
u^p(x) = d^n(x) \; ,
\quad
d^p(x) = u^n(x) \; ,
\quad
\Delta u^p(x) = \Delta d^n(x) \; ,
\quad
\Delta d^p(x) = \Delta u^n(x).
\end{equation}
The strange quarks and antiquarks are assumed to be
unpolarized ($\Delta s = \Delta \bar{s} = 0$).
This assumtion is contrary to the usual interpretation of the experimental
data which suggests $\int dx (\Delta s(x) + \Delta \bar{s}(x) \approx -0.1$.
This interpretation is, however, far from rigorous.
It depends crucially on the question whether the F/D ratio in hyperon beta
decays is affected by flavour--symmetry breaking or not, see the
discussion in \cite{BEh95}.
While the amount of symmetry breaking to be expected is strongly
model dependent, it seems to be generally accepted that an apprecially
effect can not be excluded by any rigorous undisputable argument.
The real data can be fitted excellently with $\Delta s = \Delta \bar{s} = 0$
as we shall show and as was also observed by the Durham group
\cite{TGe94}.
We are aware of theoretical considerations \cite{ERo94,ESa92,BMa94}
which predict an isospin asymmetry for the unpolarized case of the order of
10 \% for large x and the nonleading parton distribution (i.e.~$
d^p(x) \neq u^n(x)$, see also \cite{BMa94}).
For the polarized case it is common to
spin--dilution models to break isospin symmetry \cite{ASc88,DFl93}.
But as we are able to fit the data we do not see experimental
evidence for the breaking of isospin symmetry in the polarized case.

Using the parton model expressions for the structure functions:
\begin{eqnarray}
F_2(x) &=& 2xF_1(x)
\\
\nonumber \\
\label{eq10}
F_1(x) &=& \frac{1}{2} \sum_f e_f^2 q_f(x)
\\
\label{eq11}
g_1(x) &=&  \frac{1}{2} \sum_f e_f^2 \Delta q_f(x)
\end{eqnarray}
one identifies ${\cal A}_0$ and ${\cal A}_1$ with the following combinations
of $u$ and $d$ {\em valence } quark densities in order to fulfill equation
(\ref{eq10}).
\begin{equation} \label{eq12}
\frac{{\cal A}_0(x)}{x} = u_v(x) - \frac{1}{2} d_v(x),
\quad \frac{{\cal A}_1(x)}{x} = \frac{3}{2} d_v(x) \; .
\end{equation}
Note that $d_v(x)/u_v(x)$ and therefore also ${\cal A}_1(x)/{\cal A}_0(x)$
goes to zero
for $x \to 1$.
For that reason for large x the proton state is dominated by
\begin{equation} \label{eq9}
\big| p(x)\uparrow \big>  \approx \hat{u}^\dagger (x) \uparrow
\big|  I=0,I_3=0,S=0,S_3=0 \big>  \sqrt{{\cal A}_0(x)/x} \; .
\end{equation}
This formula (\ref{eq9}) reflects the intuition of the leading particle
picture, namely that for large $x$ the hidden parton will carry
the spin and isospin quantum
numbers of the proton.

As Carlitz and Kaur \cite{RCa77} discussed it is necesarry to introduce
a ``spin dilution function'' in equation (\ref{eq1}) to fulfill
the Bjorken sum rule.
For the spin dilution functions we follow the approach of reference
\cite{ASc88}
and define:
\begin{eqnarray}
F_q(I,I_3,S,S_3 = 0;x) &=&
1 - \frac{P(q\downarrow (x) \; G \uparrow (x_G),I,I_3,S,S_3 = 0)}
          {P(q \uparrow (x),I,I_3,S,S_3 = 0)}
\nonumber \\
\\
F_q(I,I_3,S,S_3 = 1;x) &=&
1 - \frac{P(q\uparrow (x) \; G \downarrow (x_G),I,I_3,S,S_3 = 1)}
          {P(q \downarrow (x),I,I_3,S,S_3 = 1)}
\nonumber
\end{eqnarray}
with $P(\dots )$ denoting the probability of finding a state with the
quantum numbers and particles in brackets.
Following the arguments in reference \cite{ASc88} the spin dilution
functions do not depend on $I$, $I_3$.
But there is a dependence on $S_3$, reflecting the $SU(6)$
symmetry breaking between the delta resonance and the nucleon.
If one assumes no dependence on $S_3$ one gets the original
Carlitz Kaur model \cite{RCa77},
which corresponds to the SU(6) value for $F/D$:
$F/D$ = 2/3.
The $F/D$ value we extract from experiment \cite{BEh95}
is however smaller $F/D = 0.49 \pm 0.08 $.

Contrary to our earlier work \cite{ASc88} we assume now that the difference
in current quark mass between $u$ and $d$ quarks does not generate
an appreciable isospin asymmetry of the spin distributions.

The remaining four independent spin dilution functions are parametrized
as follows:
\begin{eqnarray} \label{eq51}
F_q(S_3;x) &=& \frac{f_q(S_3;x) +1}{2} \; ,
\nonumber \\ \nonumber \\
f_q(S_3;x) &=& \frac{1}{1 + a(S_3)x^{-\alpha_q}(1-x)^2} \; .
\end{eqnarray}
Here the idea is the following: for small $x$ the relative
polarization, e.~g.~ of the $u^\dagger (x) \uparrow \big| 0,0,0,0 \big> $
state is proportional to
\begin{equation}
\frac{F_u(0;x) - (1 - F_u(0;x))}{F_u(0;x) + (1 - F_u(0;x))}
= 2 F_u(0;x) - 1 = f_u(0;x) \; .
\end{equation}
The rational for the choice of the dilution functions
$f_q(S_3;x)$ can be found in \cite{RCa77,ASc88}.
Basically it is just the easiest interpolation between the small--$x$
and large--$x$ limits.
This leads to the following proton state consisting of three valence
quarks and one gluon:
\begin{eqnarray} \label{eq20}
\big| p(x,x_G)\uparrow \big>  &=&
 \hat{u}^\dagger (x) \uparrow \big| 0,0,0,0 \big>
            \sqrt{F_u(0;x) {\cal A}_0(x)/x }
\nonumber \\
       &&   + 2/3 \; \hat{d}^\dagger (x) \downarrow \big| 1,1,1,1 \big>
            \sqrt{F_d(1;x) {\cal A}_1(x)/x }
\nonumber \\
        && - \sqrt{2}/3 \; \hat{d}^\dagger (x) \uparrow \big| 1,1,1,0 \big>
             \sqrt{F_d(0;x) {\cal A}_1(x)/x }
\nonumber \\
        && - \sqrt{2}/3 \; \hat{u}^\dagger (x) \downarrow
         \big| 1,0,1,1 \big>
             \sqrt{F_u(1;x) {\cal A}_1(x)/x }
\\
       &&    + 1/3 \; \hat{u}^\dagger (x) \uparrow \big| 1,0,1,0 \big>
             \sqrt{F_u(0;x) {\cal A}_1(x)/x }
%
%
\nonumber \\
        && + \hat{u}^\dagger (x) \downarrow \tilde{\hat{G}}^\dagger (x_G)
           \uparrow \big| 0,0,0,0 \big>
           \sqrt{(1-F_u(0;x)) {\cal A}_0(x)/x }
\nonumber \\ &&
          + 2/3 \; \hat{d}^\dagger (x) \uparrow \tilde{\hat{G}}^\dagger (x_G)
           \downarrow \big| 1,1,1,1 \big>
            \sqrt{(1-F_d(1;x)) {\cal A}_1(x)/x }
\nonumber \\
        && - \sqrt{2}/3 \; \hat{d}^\dagger (x) \downarrow
       \tilde{\hat{G}}^\dagger (x_G)   \uparrow \big| 1,1,1,0 \big>
             \sqrt{(1-F_d(0;x)) {\cal A}_1(x)/x }
\nonumber \\
        && - \sqrt{2}/3 \; \hat{u}^\dagger (x) \uparrow
       \tilde{\hat{G}}^\dagger (x_G)      \downarrow \big| 1,0,1,1 \big>
             \sqrt{(1-F_u(1;x)) {\cal A}_1(x)/x }
\nonumber \\ &&
           + 1/3 \; \hat{u}^\dagger (x) \downarrow
        \tilde{\hat{G}}^\dagger (x_G)     \uparrow  \big| 1,0,1,0 \big>
             \sqrt{(1-F_u(0;x)) {\cal A}_1(x)/x }
\nonumber
\end{eqnarray}
The quantum numbers of the spectator quarks in kets are $I,I_3,S,S_3$.
The part of the proton state which consists of three quarks and
a gluon depends on the fractional momentum of the active quark $x$
and on the fractional momentum of the gluon $x_G$.
$x_G$ is an additional unknown quantity.
Before the polarized gluon component was neglected in Carlitz--Kaur
type models.
As we are interested in the $Q^2$ evolution we have to take
it into account.
It seems naturally to assume $x_G \ll x$.
For explicit calculations we shall use the two choices $x_G = x/10$
and $x_G = x$, which both give very similar asymmerties $A_{1/2}$.

Note that this nucleon states give zero polarization for the
strange sea and nonzero gluon polarization.
This reflects our intuition that some ot the valence spin
is transformed to gluons.
As gluons also carry spin balanced by angular momentum our results
for $\mid \Delta \tilde{G} \mid$ are rather lower bounds than absolute
estimates for $\mid \Delta G \mid$.

Equations (\ref{eq11}), (\ref{eq12}), (\ref{eq20}) give
for the proton
\begin{equation} \label{eq30}
2xg_1^p(x) = \frac{4}{9} x u_v(x) f_u(0;x)
             - \frac{1}{27} x d_v(x)
             \left( 4 f_u(0;x) + 4 f_u(1;x) - f_d(0;x) + 2 f_d(1;x)
             \right)
\end{equation}
for the neutron \footnote{Unfortunatelly there are misprints in the
corresponding equation (17) of reference \cite{ASc88}.}
\begin{equation} \label{eq31}
2xg_1^n(x) = \frac{1}{9} x u_v(x) f_u(0,x)
             - \frac{1}{27} x d_v(x)
             \left(f_u(0;x) + f_u(1;x) - 4 f_d(0;x) + 8 f_d(1;x)
             \right)
\end{equation}
and for the gluon polarization:
\begin{eqnarray} \label{eq32}
&& x_G \Delta \tilde{G}(x_G) =
\\
&& \; = \frac{1}{2} x u_v (x) \left(1-f_u(0;x)\right) -
                \frac{1}{6} x d_v (x)
                \left(3 + f_d(0;x) - f_u(0;x) - 2 f_d(1;x) - f_u(1;x)
                \right) \nonumber
\end{eqnarray}
It is not obvious from equation (\ref{eq32}) that
$\mid \Delta \tilde{G}(x) \mid \le \mid G (x) \mid $, but we checked
it explicitly.
The parameters $\alpha_q$ are fixed by the asumption, that for
$x \to 0$ all degrees of freedom get equally populated.
More precise,
\begin{equation}
\lim _{x \to 0} f_q(S_3,x) \sim \frac{q_v (x)}{G(x)}
\sim x^{\alpha_q} \; .
\end{equation}
For the unpolarized structure functions we use the
first order (in $\alpha_s$) parametrization of
Gl\"uck, Reya, Vogt (GRV) \cite{MGl92}.
For $Q^2_0 = 10 $ GeV$^2$ they give:
$\alpha_u = 0.326$ and $\alpha_d = 0.505$.
These (new) values are substantial smaller than the ones used in
reference \cite{ASc88}.
The remaining two parameters $a(0) = a_0$ and $a(1) = a_0 \cdot a_{10}$
are choosen in a way that $g_1^{p/n}$ fulfills the Bjorken \cite{JBj66}
and Ellis Jaffe \cite{JEl74} sum rules
for $Q^2_0 = 10$ GeV$^2$. i.e.:
\begin{equation} \label{eq17}
\int_0^1 dx \left(g_1^p (x,Q^2_0=10 \; {\rm GeV}^2)
                - g_1^n (x,Q^2_0=10 \; {\rm GeV}^2)
            \right) = 0.187
\end{equation}
and
\begin{equation} \label{eq18}
\int_0^1 dx g_1^n (x,Q^2_0=10 \; {\rm GeV}^2) = -0.042 \; .
\end{equation}
This value includes the {\cal O}($\alpha_s^3$) corrections
and the {\cal O}($\alpha_s^4$) estimate for the
nonsinglet part and the {\cal O}($\alpha_s^2$)
correction and the {\cal O}($\alpha_s^3$) estimate
for the singlet part\cite{SLa91}.
It also includes the small higher twist corrections from
\cite{ESt95}.
We use $\alpha_s (10 {\rm GeV}^2) = 0.24$ from \cite{PaDa94}.
Our F and D values of
F = 0.415 , D = 0.843, with the constraint $F + D = 1.257$
are tken from our analysis \cite{BEh95}.
The resulting values (for $Q_0^2 = 10$ GeV$^2$) for quark and gluon
polarization are:
$\Delta u (Q^2 = 10 \enspace {\rm GeV}^2) = 0.746$,
$\Delta d (Q^2 = 10 \enspace {\rm GeV}^2) = -0.377$,
$\Delta s := 0$ (by construction).
This gives $a_0 = 0.225$ and $a_{10} = 0.15$ and
 $\Delta \tilde{G} (Q^2 = 10 \enspace {\rm GeV}^2) = 0.315$.
Note that in our model $\Delta u (Q_0^2= 10 \enspace {\rm GeV}^2 )
+ \Delta d (Q_0^2= 10 \enspace {\rm GeV}^2 )
+ 2\Delta \tilde{G}(Q_0^2= 10 \enspace {\rm GeV}^2 ) = 1$, as some
of the quarks spin is transfered to the gluons, but no angular
momentum is generated.
Figure \ref{fig1} shows the polarized parton densities
$\Delta u$, $\Delta d$, $\Delta \tilde{G}$, for $Q_0^2 = 10$ GeV$^2$.
The behavior for $x \to 0$ and $x \to 1$ of the distributions is
as follows:
\begin{eqnarray}
\label{eq17a}
\lim_{x \to 1} x \Delta u(x) &=& xu_v(x)
\\ \nonumber
\\ \label{eq17b}
\lim_{x \to 1} x \Delta d(x) &=& - \frac{1}{3}xd_v(x)
\\ \nonumber
\\ \label{eq17c}
\lim_{x_G \to 1} x_G \Delta \tilde{G}(x_G) &=& \frac{a_0}{2} xu_v(x) (1-x)^2
\\ \nonumber
\\ \label{eq17d}
\lim_{x \to 0} x \Delta u(x) &=& x^{2\alpha_u} \approx x^{0.652}
\\ \nonumber
\\ \label{eq17e}
\lim_{x \to 0} x \Delta d(x) &=& x^{2\alpha_d} \approx x^{1.010}
\\ \nonumber
\\ \label{eq17f}
\lim_{x_G \to 0} x_G \Delta \tilde{G}(x_G) &=& \frac{1}{2} x u_v(x)
\approx x^{0.326}
\end{eqnarray}
$\Delta u$ shows the behaviour predicted by the
quark counting rules \cite{SBr94}, however $\Delta d $
does not.
In the framework of the spin dilution model it is impossible
to get a counting rule like behaviour for both $\Delta u$ and
$\Delta d$ if one assumes
$\lim_{x \to 1} F_q(S_3,x) = 1$\footnote{This behaviour is natural
if one has a spin dilution in mind, thinking of the $F_q$'s as
coefficients of a general Melosh (spin) rotation \cite{FCl74,HMe74}
this constraint does not necessarily exist.}.
A counting rule like behaviour of $\Delta u$
is essential in order to obtain the leading particle picture.
The precise behavior of
$\Delta d (x)$ is unimportant for the leading particle picture as
$\Delta d (x)$ is limited by $\mid \Delta d (x) \mid \le d(x)$.
Change the behaviour of $\Delta d$ ad hoc in order to
fulfill the quark counting rules leads to results which contradict
existing data\footnote{Note the papers of Artru \cite{XAr80},
where he derives an alternative to the dimensional counting
rules with different predictions.}.

It was argued \cite{LFr89} that important new information
can be obtained from the analysis of semi--inclusive processes.
For the $\pi$--asymmetry $A_\pi$ defined as
\begin{equation}
A_\pi (x) = \frac{N_{\uparrow \downarrow}^{\pi^+ - \pi^-}
                  - N_{\uparrow \uparrow}^{\pi^+ - \pi^-}}
                 {N_{\uparrow \downarrow}^{\pi^+ - \pi^-}
                  + N_{\uparrow \uparrow}^{\pi^+ - \pi^-}}
\end{equation}
the pion asymmetry for proton, deuterium and $~^3$He is given by:
\begin{eqnarray}
A_\pi^p (x) &=& \frac{4\Delta u^v (x) - \Delta d^v (x)}
                     {4 u^v (x) -  d^v (x)} \; ,
\\
A_\pi^d (x) &=& \frac{\Delta u^v (x) + \Delta d^v (x)}
                     {u^v (x) +  d^v (x)}  \; ,
\\
A_\pi^{~^3He} (x) &=& \frac{4\Delta d^v (x) - \Delta u^v (x)}
                     {7 u^v (x) +  2 d^v (x)} \; .
\end{eqnarray}
Figure \ref{fig1b} shows our model prediction for this asymmetries.
%
\section{A modified spin dilution model \label{sec2a}}
If one does not insist to get a counting rule like behaviour
for $\Delta u$ it is also possible to construct a proton state
e.~g.~ according to:
\begin{eqnarray} \label{eq100}
\big| p(x)\uparrow \big>  &=&
\left( \hat{u}^\dagger (x) \uparrow \big| 0,0,0,0 \big>
          + 1/3 \; \hat{u}^\dagger (x) \uparrow \big| 1,0,1,0 \big>
           \right.
\nonumber \\
&&  \left. - \sqrt{2}/3 \; \hat{u}^\dagger (x) \downarrow \big| 1,0,1,1
\big>
\right)\sqrt{\tilde{\cal A}_0(x)/x}
\nonumber \\
         && + \left( 2/3 \; \hat{d}^\dagger (x) \downarrow
         \big| 1,1,1,1 \big>
       - \sqrt{2}/3 \; \hat{d}^\dagger (x) \uparrow
        \big| 1,0,1,0 \big>  \right)
        \sqrt{\tilde{\cal A}_1(x)/x}
\end{eqnarray}
instead of equation (\ref{eq1}), thus giving up approximate SU(6)
symmetry.
In equation (\ref{eq100}) $\tilde{\cal A}_0(x)$ is proportional to
$u_v(x)$ and $\tilde{\cal A}_1$ is proportional to $d_v(x)$.
Therefore the probability to find a quark of flavor $q$ with definite spin
will only depend on the unpolarized distribution function of the same
flavor $q(x)$, which seems to be much more natural compared to
equation (\ref{eq1}), were e.g. the probability to find a $u$ quark
with spin down depends only on $d_v(x)$, but not on $u_v(x)$.
(This mixing of up and down distribution functions obviously reflects
the SU(6) symmetry.)

Introducing the spin dilution fuctions of equation (\ref{eq51})
gives:
\begin{eqnarray} \label{eq101}
\big| p(x,x_G)\uparrow \big>  &=&
 \left( \left( \hat{u}^\dagger (x)
                            \uparrow \big| 0,0,0,0 \big>
               \right. \right.
\nonumber \\
         && + \left. 1/3 \; \hat{u}^\dagger (x) \uparrow \big| 1,0,1,0 \big>
              \right) \sqrt{F_u(0;x)}
\nonumber \\
&& - \left. \sqrt{2}/3 \; \hat{u}^\dagger (x) \downarrow \big| 1,0,1,1 \big>
     \sqrt{F_u(1;x)} \right)\sqrt{u(x)3/4}
\nonumber \\
         && + \left( 2/3 \; \hat{d}^\dagger (x) \downarrow
         \big| 1,1,1,1 \big>  \sqrt{F_d(0;x)}\right.
\nonumber \\
   && - \left. \sqrt{2}/3 \; \hat{d}^\dagger (x) \uparrow
        \big| 1,0,1,0 \big>  \sqrt{F_d(1;x)} \right)
         \sqrt{d(x)3/2}
\nonumber \\
%
%
&& + \left( \left( \hat{u}^\dagger (x) \downarrow
              \tilde{\hat{G}}^\dagger (x_G) \uparrow
              \big| 0,0,0,0 \big>
               \right. \right.
\nonumber \\
         && + \left. 1/3 \; \hat{u}^\dagger (x) \downarrow
             \tilde{\hat{G}}^\dagger (x_G) \uparrow \big| 1,0,1,0 \big>
              \right) \sqrt{1 - F_u(0;x)}
\nonumber \\
&& - \left. \sqrt{2}/3 \; \hat{u}^\dagger (x) \uparrow
      \tilde{\hat{G}}^\dagger (x_G) \downarrow \big| 1,0,1,1 \big>
     \sqrt{1 - F_u(1;x)}
     \right)\sqrt{u(x)3/4}
\nonumber \\
         && + \left( 2/3 \; \hat{d}^\dagger (x) \uparrow
         \tilde{\hat{G}}^\dagger (x_G) \downarrow \big| 1,1,1,1 \big>
         \sqrt{1- F_d(0;x)} \right.
\nonumber \\
   && - \left. \sqrt{2}/3 \; \hat{d}^\dagger (x) \downarrow
        \tilde{\hat{G}}^\dagger (x_G) \uparrow
        \big| 1,0,1,0 \big>  \sqrt{1 - F_d(1;x)} \right)
        \sqrt{d(x)3/2}
\end{eqnarray}
This proton state gives:
\begin{eqnarray} \label{eq102}
\Delta u(x) &=& \frac{1}{6} u_v(x) \left( 5 f_u(0;x) - f_u(1;x)\right)
\nonumber \\
\Delta d(x) &=& \frac{1}{3} d_v(x) \left( f_d(0;x) - 2f_d(1;x)\right)
\\
\Delta \tilde{G}(x_G) &=& \frac{1}{12} u_v(x)
               \left( 4 - 5 f_u(0;x) + f_u(1;x)\right)
              + \frac{1}{6} d_v(x) \left( 2f_d(1;x) - f_d(0;x) - 1 \right)
\nonumber
\end{eqnarray}
With $a_0 = 0.233$ and $a_{10} = 0.145$ these distributions fulfill equations
(\ref{eq17}) and (\ref{eq18}). Their behaviour for $x \to 0,1$ is:
\begin{eqnarray} \label{eq110a}
\lim_{x \to 1} x \Delta u(x) &=& \frac{2}{3}xu_v(x)
\\ \nonumber \\ \label{eq110b}
\lim_{x \to 1} x \Delta d(x) &=& - \frac{1}{3}xd_v(x)
\\ \nonumber \\ \label{eq110d}
\lim_{x \to 0} x \Delta u(x) &=& x^{2\alpha_u} \approx x^{0.652}
\\ \nonumber \\ \label{eq110e}
\lim_{x \to 0} x \Delta d(x) &=& x^{2\alpha_d} \approx x^{1.010}
\\ \nonumber \\ \label{eq110f}
\lim_{x_G \to 0} x_G \Delta G(x_G) &=& \frac{1}{3} x u_v(x) \approx x^{0.326}
\end{eqnarray}
Figure \ref{fig1a} shows the resulting distributions.
The main difference between the two models is the
behaviour of $\Delta u$ for large x. The model in this section gives
$\lim_{x \to 1} = 2/3$ for the asymmetry $A_1$ for proton and neutron.
This contradicts the leading particle picture.
However we are able to
fit the experimental results with this model, as will be seen in the next
section.
As a consequence it seems to be extremly important to get more
accurate data at high $x$ in order to check the validity of the
leading particle picture.

%
\section{$Q^2$ evolution and comparison with data \label{sec3}}
Using the first order GLAP equations \cite{GAl77} with
$\Delta s (x,Q^2) = 0$, $ n_f = 2$, but no additional
asumptions we numerically\footnote{We do not have any problems with the
stability or accuracy of the numerical solutions for any values of $Q^2$
and $x$.}
generate from equations (\ref{eq30}) - (\ref{eq32}), (\ref{eq102})
$g_1$ for the experimental values of $Q^2$ by evolution from
$Q_0^2 = 10$ GeV$^2$.
Figures \ref{fig2}, \ref{fig2a}, \ref{fig3}, \ref{fig3a} show a comparison of
the virtual photon asymmetry $A_1$\cite{KRi89,MAn94}:
\begin{equation} \label{eq39}
A_1^{p/n} = \frac{g_1^{p/n}(x,Q^2) - 4g_2^{p/n}(x,Q^2)M^2x^2/Q^2}
                 {F_1(x,Q^2)}
\end{equation}
with experimental data \cite{E130,JAs88,PAn93,BAd93,DAd94,KAb95}.
$M$ denotes the mass of the proton.
As we neglegt the effect of twist 4 and higher operatos as well as
${\cal O}(\alpha_s^2)$ corrections we insert the parton model
expression (\ref{eq10}) for $F_1$ (with sea quarks included)
into equation (\ref{eq39}), which we take from \cite{MGl92}.
Instead we could also substitute $F_1$ by
the parton model expression for $F_2$ (\ref{eq11})
and $R$: $F_1 = 2x F_2 (1+R)$.
Using the standard parametrization
for $R$ \cite{LWh90}, which includes higher twist and higher
order radiative corrections, this would lead to different results for
$A_1$. Most notably for large $x$, $R(x,Q^2) > 0$ and therefore
for this choice of asymmetry $A_1(x,Q^2) < 1$ for $x \to 1$ even if the
parton densities show leading particle behaviour.
As this is hard to conciliate with physical intuition we use $F_1$(parton)
instead.

As there exists no experimental data for $g_2(x,Q^2)$ we considered two
rather different assumtions for it.
\begin{equation} \label{eq40}
g_2(x,Q^2) = 0 \; ,
\end{equation}
which is the simple parton model prediction and implies that the twist three
part of $g_2$ is as large as the twist two part and exactly cancels it.
And
\begin{equation} \label{eq41}
g_2(x,Q^2) = g^{\rm tw \; 2}_2(x,Q^2) = -g_1(x,Q^2) +
    \int_x^1 \frac{dy}{y} g_1(y,Q^2) \; ,
\end{equation}
which means that the twist three part of $g_2$ is zero.
We expect the physical reality to lie somewhere
inbetween these two extreme scenarios.
\\ \\
Figures \ref{fig4} - \ref{fig7} show the resulting predictions
for $A^{p/n}_2(x,Q^2)$
\begin{equation}
A^{p/n}_2(x,Q^2) \sqrt{Q^2}/M = 2x \frac{g_1^{p/n}(x,Q^2) + g_2^{p/n}(x,Q^2)}
                                        {F_1^{p/n}(x,Q^2)}
\end{equation}
with the asumptions (\ref{eq40}) and (\ref{eq41}) for $g_2$.
%
\section{\label{sec4} Conclusion}
We presented two variants of spin dilution models that describe the
data. The models give predictions for $\Delta u(x)$ and $\Delta d(x)$ and
a kind of lower bound for $\Delta G(x) $.
They fit the experimental data very well.
These fits fulfill the Bjorken and Ellis--Jaffe sum rules
with appropriate choosen F and D values.
Therefore we do not see experimental evidence for isospin violation
(as originally advocated in \cite{ASc88}) or for a
strong strange sea polarization.
As we could not fit data with a $\Delta d(x)$ that behaves like the
counting rule prediction we doubt the correctness of the counting rules for
$\Delta d$. Furthermore it seems us by now impossible to decide from data,
whether the counting rule prediction (and the leading particle picture!)
for $\Delta u$ is right.
Our calculation gives a small, but systematic, dependence
of $A_1$ and $\sqrt{Q^2} \cdot A_2$ on $Q^2$.

This work was supported in part by DFG (G. Hess Program),
BMFT, GSI and Cusanuswerk.
\enspace A.S. thanks also the MPI f\"ur Kernphysik in Heidelberg for its
hospitality.

\newpage

{\LARGE \bf Figure Caption}
\rm
\setcaptype{figure}
\caption{solid line $x\Delta u(x,Q^2_0)$, dotted line $x\Delta d(x,Q^2_0)$,
dashed line $x\Delta \tilde{G}(x=x_G,Q^2_0)$ for $Q^2_0 = 30$ GeV$^2$,
for the model of section 2.}
\label{fig1}
\caption{Pion--asymmetries for proton (solid line), deuterium (dotted line)
and $He^3$ (dashed line).}
\label{fig1b}
\caption{solid line $x\Delta u(x,Q^2_0)$, dotted line $x\Delta d(x,Q^2_0)$,
dashed line $x\Delta \tilde{G}(x=x_G,Q^2_0)$ for $Q^2_0 = 30$ GeV$^2$,
for the model of section  3.}
\label{fig1a}
\caption{ $A_1^p(x,Q^2)$:
solid line $Q^2 = 1 $ GeV$^2$,
large dashed line $Q^2 = 4 $ GeV$^2$,
dashed line $Q^2 = 10 $ GeV$^2$,
dotted line $Q^2 = 40 $ GeV$^2$, for the model of section
2.
Diamonds SMC, circles EMC, Squares E130 measurements.}
\label{fig2}
\caption{ $A_1^p(x,Q^2)$:
solid line $Q^2 = 1 $ GeV$^2$,
large dashed line $Q^2 = 4 $ GeV$^2$,
dashed line $Q^2 = 10 $ GeV$^2$,
dotted line $Q^2 = 40 $ GeV$^2$, for the model of section
3.
Diamonds SMC, circles EMC, Squares E130 measurements.}
\label{fig2a}
\caption{ $A_1^n(x,Q^2)$:
solid line $Q^2 = 1 $ GeV$^2$,
large dashed line $Q^2 = 4 $ GeV$^2$, for the model of section
2.
dashed line $Q^2 = 10 $ GeV$^2$,
dotted line $Q^2 = 40 $ GeV$^2$,
Diamonds E142 measurement.}
\label{fig3}
\caption{ $A_1^n(x,Q^2)$:
solid line $Q^2 = 1 $ GeV$^2$,
large dashed line $Q^2 = 4 $ GeV$^2$, for the model of section
3.
dashed line $Q^2 = 10 $ GeV$^2$,
dotted line $Q^2 = 40 $ GeV$^2$,
Diamonds E142 measurement.}
\label{fig3a}
\caption{$A_2^p(x,Q^2)$ for $g_2 =0$ and for the model of section
2.
Solid line $Q^2 = 1 $ GeV$^2$,
large dashed line $Q^2 = 4 $ GeV$^2$,
dashed line $Q^2 = 10 $ GeV$^2$,
dotted line $Q^2 = 40 $ GeV$^2$.}
\label{fig4}
\caption{$A_2^p(x,Q^2)$ for $g_2 = g_2^{\rm tw 2}$ and for the model of
section
2.
Solid line $Q^2 = 1 $
GeV$^2$,
large dashed line $Q^2 = 4 $ GeV$^2$,
dashed line $Q^2 = 10 $ GeV$^2$,
dotted line $Q^2 = 40 $ GeV$^2$.}
\label{fig5}
\caption{$A_2^n(x,Q^2)$ for $g_2 =0$ and for the model of section
2.
Solid line $Q^2 = 1 $
GeV$^2$,
large dashed line $Q^2 = 4 $ GeV$^2$,
dashed line $Q^2 = 10 $ GeV$^2$,
dotted line $Q^2 = 40 $ GeV$^2$.}
\label{fig6}
\caption{$A_2^n(x,Q^2)$ for $g_2 = g_2^{\rm tw 2}$ and for the model of
section
2.
Solid line $Q^2 = 1 $
GeV$^2$,
large dashed line $Q^2 = 4 $ GeV$^2$,
dashed line $Q^2 = 10 $ GeV$^2$,
dotted line $Q^2 = 40 $ GeV$^2$.}
\label{fig7}

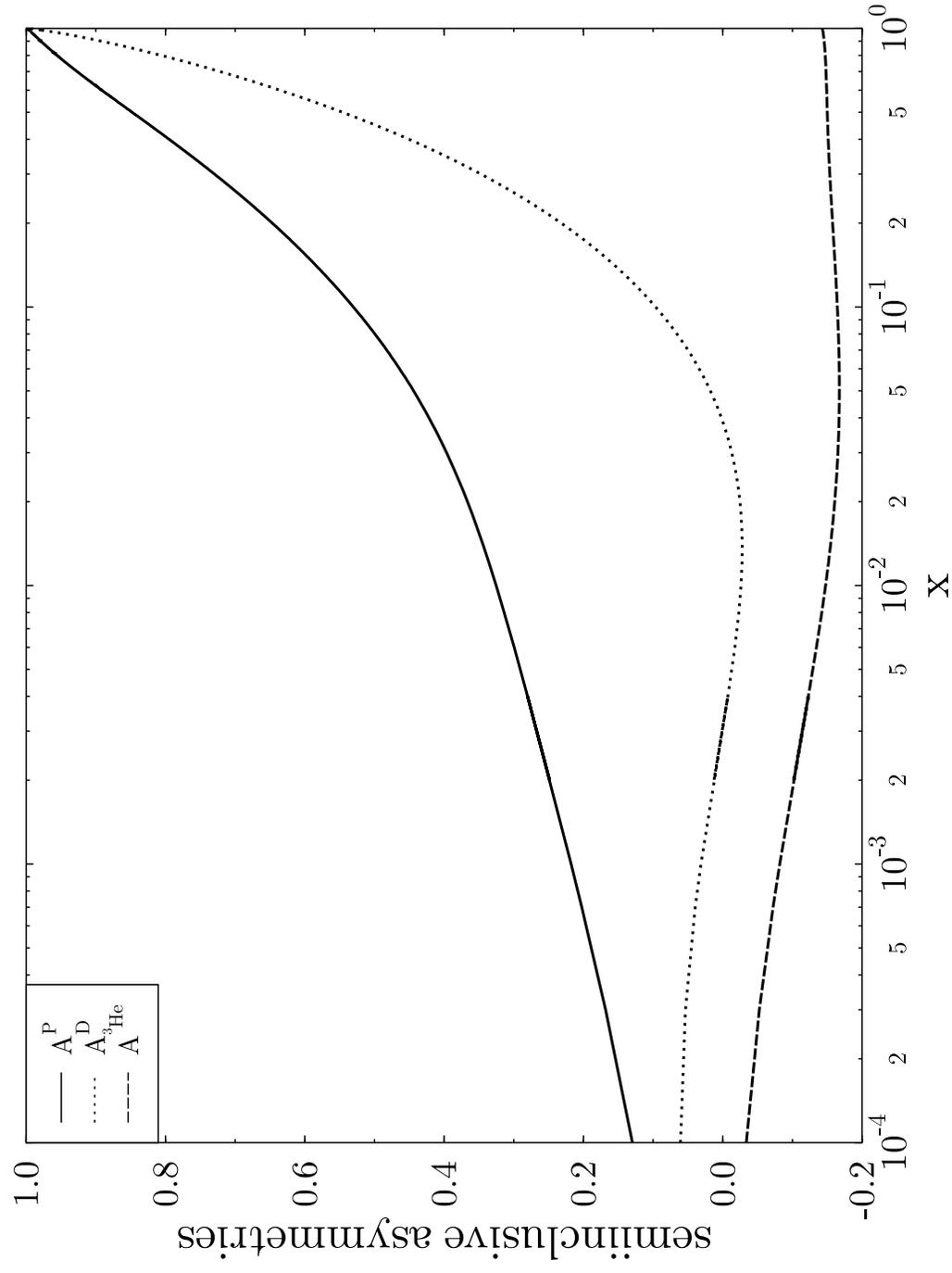
\end{document}